# Luminescence properties and phase transformation of broadband NIR emitting $A_2(WO_4)_3$:$Cr^{3+}$ (A=$Al^{3+}$, $Sc^{3+}$) phosphors toward NIR spectroscopy applications

Shuai Yang[1], Yongjie Wang[1,2,*], Guotao Xiang[1], Sha Jiang[1], Li Li[1], Faling Ling[1], Huanhuan Hu[1], Yuanyuan Zhang[3,4], Xianju Zhou[1,2*], Andrzej Suchocki[5]

[1]*College of Sciences, Chongqing University of Posts and Telecommunications, Chongqing 400065, China*
[2]*Chongqing Key Laboratory of Photoelectronic Information Sensing and Transmitting Technology, Chongqing University of Posts and Telecommunications, Chongqing 400065, China*
[3]*School of Materials Sciences and Technology, Dongguan University of Technology, Dongguan 523808, China.*
[4]*School of Materials Science and Engineering, Xi'an Jiaotong University, Xi'an 710049, China.*
[5]*Institute of Physics, Polish Academy of Sciences, Al. Lotników 32/46, 02-668 Warsaw, Poland*

**Abstract**

The synthesis, structural, and luminescence properties have been carried out for $Cr^{3+}$-activated $Al_2(WO_4)_3$ (AWO) and $Sc_2(WO_4)_3$ (SWO) phosphors for application in pc-NIR LED. Upon blue excitation, these compounds are capable of exhibiting broadband NIR emission stems primarily from $^4T_2 \rightarrow {}^4A_2$ transition in the range of 670-1200 nm (maxima ~808 nm, FWHM ~140 nm) for AWO:Cr and of 700-1300 nm (maxima ~870 nm, FWHM ~164 nm) for SWO:Cr. The significant shift of NIR emission is attributed to the substitution of $AlO_6$ with larger $ScO_6$ octahedrons. To gain insight into the luminescence the crystal field strength, Racah parameters, nephelauxetic effect, and electron-phonon coupling have been analyzed based on spectroscopic results. The electron-phonon coupling parameter *S* for SWO:Cr was determined to be 11.5, twice as large as that for AWO:Cr, which is in accordance with its strong thermal quenching. The abrupt changes occurring at 275 K in temperature-dependent luminescence spectra and decay lifetime of AWO:Cr is associated with temperature-driven phase transformation from low-temperature monoclinic to high-temperature orthorhombic phase. Pressure induced amorphization of AWO:Cr at pressures higher than 25 kbar was confirmed by employing high pressure evolution of Raman spectra. A high-power NIR pc-LED, fabricated by coating AWO:0.04Cr on a commercial 470 nm LED chip, shows good performance with an output power of 17.1 mW driven by a current of 320 mA, revealing potential application of studied materials for NIR light source.

1. Introduction

NIR spectroscopy offers numerous advantages over traditional methods, such as fast detection, nondestructive, and noninvasive examinations and non-contact technique to name a few [1]. These advantages of NIR spectroscopy methods have attracted great interest in analytical techniques, including agriculture operations, health monitoring, pharmaceutical science, and food safety detection due to its minimum damage to specimens, and deep sampling [2,3,4,5]. However, the ongoing miniaturization and intelligence of integrated spectrometers are challenging because traditional NIR light sources, such as tungsten halogen lamps, globars, and tunable lasers have large volumes and trend to high energy consumption and low efficiency, which restricts their practical applicability[6]. An alternative candidate for broadband NIR light sources, currently under development, is phosphor-converted NIR light-emitting diodes (pc-NIR LEDs), which are fabricated by the combination of the NIR phosphors with the blue LED chips for excitation and manifest remarkable advantages over traditional light sources, such as smaller size, spectral stability, low cost, and customized tunable broadband spectral distribution [7]. Furthermore, the figures of merit (such as high efficiency, long lifetime, and compactness) inherited from white LEDs [8] make pc-NIR LEDs suitable as light sources for miniature or portable handheld NIR spectrometers. Thus, the design and development of blue light excitable broadband NIR phosphors with high efficiency and tunable emission are the important goals for optimizing pc-NIR LEDs performance and extend the range of applications for miniaturized spectrometers.

$Cr^{3+}$ ion, which belongs to $[Ar]3d^3$ electronic configuration, is such activator of great interest for broadband tunable solid state laser gain materials in the NIR range when experienced in a weak crystal field (CF) [9][10][11]. Driven by the potential application in broadband NIR light sources based on pc-NIR LEDs, $Cr^{3+}$-activated broadband NIR phosphors excited by blue light, which matches well with the highly efficient blue LED chips, have received considerable attention in recent years [12,13,14,15]. Particularly in Cr-doped garnet-type phosphors, such as

Ca$_3$Sc$_2$Si$_3$O$_{12}$ [16], Y$_{3-x}$Ca$_x$Al$_{5-x}$Si$_x$O$_{12}$ [17], X$_3$Sc$_2$Ga$_3$O$_{12}$ (X=Lu,Y,Gd,La) [18], Ca$_2$LuX$_2$Al$_3$O$_{12}$ (X=Zr,Hf) [19,20,21], Gd$_3$Sc$_2$Ga$_3$O$_{12}$ [22] and to name a few [23,24,25,26], with high thermal quenching temperatures and internal quantum efficiency (IQE) from 58% up to ~90% are reported. Except for typical Cr$^{3+}$-activated NIR-emitting phosphors, other activators such as Eu$^{2+}$ [27][28], Sb$^{3+}$ [29] and Fe$^{3+}$ [30] doped NIR-emitting phosphors with high efficiency and tunable spectral range have also been reported.

Following our ongoing interest in discovering alternative Cr$^{3+}$-based NIR phosphors suitable for pc-NIR LED, we turned our attention to tungstates, i.e., X$_2$(WO$_4$)$_3$ (X=Al and Sc) as for host materials which may permit efficient broadband NIR emission. X$_2$(WO$_4$)$_3$ belong to A$_2$(MO$_4$)$_3$ (A=Al, Sc, Y, etc., and M=W, Mo) family materials, which are of great interest for electronic components, printed circuit boards, optical substrates, due to their negative thermal expansion [31,32] and unusual high trivalent ion electrical conduction [33]. Moreover, the spectroscopic study of tungstates and molybdates single crystals given their potential application as host lattices for tunable Cr$^{3+}$ solid state laser has examined that Cr$^{3+}$ experiences a weak CF, thus exhibiting broadband NIR luminescence originating to the spin-allowed $^4$T$_2$ → $^4$A$_2$ transition upon excitation by blue light [34,35,36]. Due to the acentric position of Cr$^{3+}$ substituting on the Al/Sc sites with point group C$_1$, these materials exhibit strong absorption and emission cross-sections and belong to the class of high-gain Cr materials. In this work, a systematic study on the synthesis, structural and luminescence properties have been investigated for Cr$^{3+}$-doped Al$_2$(WO$_4$)$_3$ and Sc$_2$(WO$_4$)$_3$ phosphors for application in pc-NIR LED. Both compounds permit broadband NIR emission stems primarily from $^4$T$_2$→$^4$A$_2$ transition under blue excitation. The crystal field parameters, nephelauxetic effect, and electron-phonon coupling effect of octahedrally coordinated Cr$^{3+}$ have been analyzed based on spectroscopic results. Additionally, a temperature-induced structural phase transition and pressure-induced amorphization for AWO:Cr compound have been identified using temperature-dependent PL spectra and high pressure Raman spectra, respectively. Finally, the performance of a pc-NIR LED device based on a 470 nm LED chip fabricated by coating AWO:0.04Cr phosphor was evaluated for the assignment of as-prepared material for NIR light sources.

2. **Experimental methods**
  2.1 **Synthesis**

Al$_{2-x}$Cr$_x$(WO$_4$)$_3$ and Sc$_{2-x}$Cr$_x$(WO$_4$)$_3$ (nominal compositions x=0, 0.01, 0.02, 0.04, 0.06 and 0.08) phosphors, here abbreviated as AWO:xCr and SWO:xCr, were readily synthesized by high-temperature solid-state reaction method in the air atmosphere. All chemicals were commercially purchased and used as received without further purification. In the preparation, Al$_2$O$_3$ (99.99%), Sc$_2$O$_3$ (99.9%), WO$_3$ (99.9%) and Cr$_2$O$_3$ (99.99%) were used as starting materials according to the stoichiometric ratio. Each mixture was dispersed in a small amount of ethanol and adequately ground in an agate mortar. The mixtures were pressed into a small pellet with diameter of ~ 1 cm, then transferred to open alumina crucibles and heated first at 900 °C for AWO:xCr or 1000 °C for SWO:xCr, held for 10 h, and then cooled to room temperature. The samples then were well ground and pressed into the pellet again. Finally, they were calcined at 1100 °C for AWO:xCr or 1200 °C for SWO:xCr, and held for 24 h with an intermediate grinding. After the samples were cooled to room temperature, they were ground to fine powders for subsequent characterizations.

## 2.2 Characterization

The XRD pattern was collected using a Bruker D2 Phaser diffractometer (Billerica, MA, USA) using Cu radiation (λ=1.54 Å) source and operated at 10 mA and 30 kV. The UV-Vis-NIR diffuse reflectance spectroscopy (DRS) was measured using Agilent Cary 5000 spectrophotometer with the use of BaSO$_4$ as a reflectance standard. The photoluminescence excitation (PLE) and photoluminescence (PL) spectra were recorded using a spectrofluorometer (Edinburgh Instruments, FLS-1000) equipped with a 450 W xenon lamp for excitation and a Hamamatsu's R928 PMT-900 detector (185-900 nm) and NIR-PMT (up to 1700 nm). Temperature dependent stead-state and life-time photoluminescence measurements were performed using an ARS cryostat (model CS204AE-FMX-1AL) integrated with an FLS1000 spectrofluorometer and a Lake Shore 335 temperature controller. The spectra of pc-NIR LED device were recorded using an OHSP-350M LED Fast-Scan Spectrophotometer covering a range of 350-1050 nm (Hangzhou Hopoo Light&Color Technology Co., Ltd. ).

Raman measurements were recorded using a confocal Raman microscope (MonoVista CRS+) with Cobolt 532 nm solid state laser. The laser was focused using a 5× Mitutoyo Objective from TM Series with a numerical aperture of 0.10. The high-pressure Raman measurements were performed using a low-temperature diamond anvil cell (CryoDAC LT, easyLab Technologies

Ltd.). Argon was used as a pressure transmitting medium. The powder samples were loaded into the cell along with a small ruby crystal. The R1-line ruby luminescence was used for pressure calibration.

## 3. Results and discussion

### 3.1 Crystal structure and phase identification

Both $Al_2(WO_4)_3$ and $Sc_2(WO_4)_3$ crystallize in the orthorhombic with $Pbcn$ ($D_{2h}^{14}$) space group, adopting a three-dimensional framework of corner-linked Al/ScO$_6$ octahedra and WO$_4$ tetrahedra [37], as confirmed by x-ray diffraction ( Fig. 1a,b). All observed peaks can be well indexed to the standard data of $Al_2(WO_4)_3$ (COD-1001563), confirming that desired compounds were readily synthesized through solid-state reaction. In the structure, $Cr^{3+}$ ions intentionally substitute for Al/Sc ions in distorted octahedra with site symmetry $C_1$. As seen in zoomed views in Figure 1, the main diffraction peak shifts to smaller angles with increasing $Cr^{3+}$ concentration for AWO:xCr, while the change for SWO:xCr towards larger angles. The main reason for the opposite effect is that the ionic radius of $Cr^{3+}$ ions (0.615 Å, CN=6) is smaller than $Al^{3+}$ ions (0.535 Å, CN=6), but larger than $Sc^{3+}$ ions (0.745 Å, CN=6). Thus the substitution of $Cr^{3+}$ induces the increase/decrease in the lattice parameters that is reflected by the shift of diffraction peaks. As has been known, the crystal field strength and its energy levels of $Cr^{3+}$ differ much from its coordinate environment. To estimate the lattice parameters, XRD patterns of AWO:0.4Cr and SWO:0.6Cr phosphors were refined using the Rietveld method with Fullprof software. The refinements were carried out in the orthorhombic structure with space group $Pbcn$. The crystallographic data as determined from Rietveld refinement are presented in Table 1. The average value of Al-O and Sc-O bond lengths are thus determined to be 1.887 and 2.078 Å.

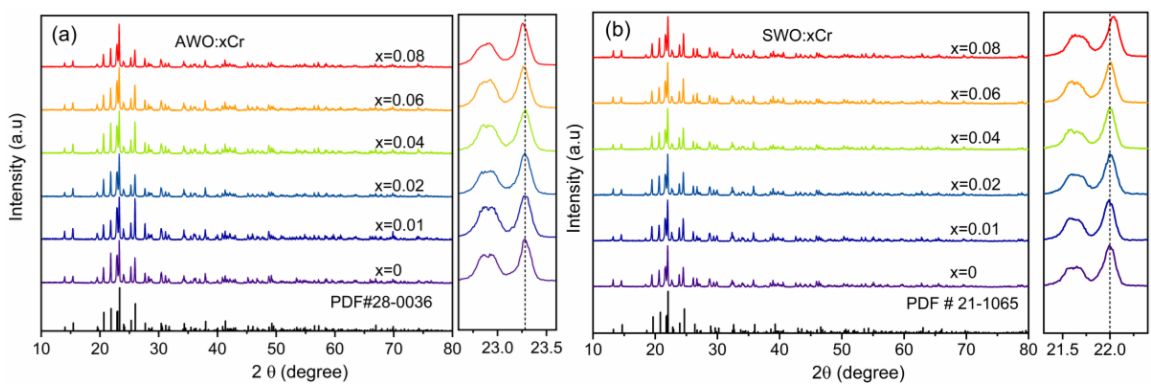

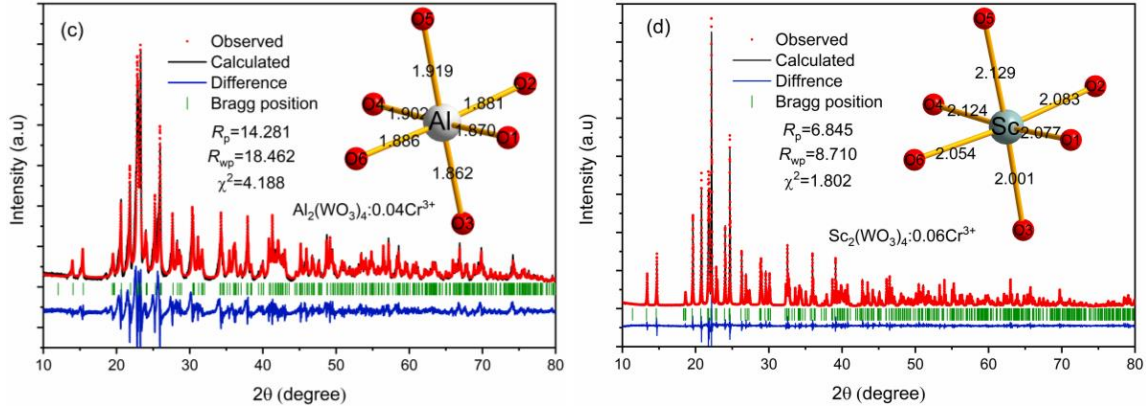

**Fig. 1.** XRD patterns and enlarged view of the main XRD peak of (a) AWO:xCr$^{3+}$ and (b) SWO:xCr$^{3+}$ samples. Rietveld refinement of XRD collected from (c) AWO:0.04Cr$^{3+}$ and (d) SWO:0.06Cr$^{3+}$; the inserts showing coordination and bond lengths of Al/Sc atoms in octahedral sites. The bond lengths are given in units of Å.

**Table 1.** Lattice parameters (labelled as a, b and c), volume of unit cell (V) and average distance between central metal ion (Al/Sc$^{3+}$) and ligands (O$^{2-}$) in the AWO:0.04Cr and SWO:0.06Cr compounds.

| Compounds | AWO:0.04Cr | SWO:0.06Cr |
| --- | --- | --- |
| Lattice parameters (Å) | a=9.1361(2), b=12.5935(3), c=9.05610(2) | a=9.66633(1) b=13.3088(4) Å c=9.5715(4) |
| volume of unite cell (Å$^3$) | V=1041.96(1) | V=1230.96(4) |
| Average bond length of Al/Sc-O | 1.8787 | 2.078 |

### 3.2 Luminescence properties of AWO:xCr$^{3+}$ and SWO:xCr$^{3+}$ phosphors

Fig. 2a and b present typical DRS, PL, and PLE spectra of AWO:xCr$^{3+}$ and SWO:xCr$^{3+}$ phosphors, respectively. Since the crystal field strength 10$Dq$ is inversely proportional to the fifth-power of metal to ligand distance $R$, the vibration of $R$ will cause significant changes in the crystal field splitting of Cr$^{3+}$. Thus a red-shift and broadening PL spectrum of Cr$^{3+}$ is expected when Al$^{3+}$ was substituted by Sc$^{3+}$ with a larger ionic radius. As seen, the broad emission band (650-1300 nm) peaking at 808 for AWO:0.04Cr$^{3+}$ or 868 nm for SWO:0.06Cr$^{3+}$ is ascribed to spin-allowed $^4T_2 \rightarrow {}^4A_2$ transition of Cr$^{3+}$ ion. A small fraction of R-line luminescence from spin-forbidden $^2E \rightarrow {}^4A_2$ transition at 723 nm is observed in AWO:xCr. In DRS and PLE spectra for AWO:0.04Cr$^{3+}$, the broad bands with barycenter at 300, 468, and 653 nm are associated with $^4A_2(^4F) \rightarrow {}^4T_1(^4P)$, $^4A_2(^4F) \rightarrow {}^4T_1(^4F)$ and $^4A_2(4F) \rightarrow {}^4T_2(4F)$ transitions, while the sharp peak 723 nm is attributed to $^4A_2 \rightarrow {}^2E$ electronic absorption. In addition, a shoulder (505 nm) assigned to spin-forbidden $^4A_2(^4F) \rightarrow {}^2T_2$ transition can be resolved. An extra weak band at 350 nm, as well as previously observed by Petermann et al [34] in the AWO single crystal, is not yet identified. By comparison with the spectrum of AWO:xCr, the sharp line at 725 nm in SWO:xCr (Fig.3b) is unambiguously assigned to $^4A_2 \rightarrow {}^2E$ transition because the $^2E$ does not involve any change in electronic configuration and is therefore nearly independent of crystal field strength. And for these $^4T_1(^4P)$ (316 nm), $^4T_1(^4F)$ (484 nm), and $^4T_2$ (694 nm) absorption bands, and sharp $^2T_2$ (508 nm) of $Cr^{3+}$ are observed as well. As can be seen, all absorption bands of $Cr^{3+}$ in SWO are shifted to longer wavelengths, following with the large $Sc^{3+}$ site. Furthermore, a pronounced Fano-antiresonance feature is observed at 680 nm for AWO:Cr or at 675 nm for SWO:Cr. The effect results from the interaction via spin-orbit coupling of the sharp electronic $^2T_1(G)$ level when superposing the vibrationally broadened $^4T_2(F)$ band as a 'quasicontinuum', thus leading to characteristically asymmetric line shapes with a more or less deep dip on the high-energy side [38]. One should mention that a subtractive peak due to Fano-antiresonance between $^4T_2$ and $^2E$ state can be resolved when $Cr^{3+}$ is in a weak octahedral crystal field [39,40,41,42]. In the case of present results, however, an apparent sharp peak originating from $^4A_2 \rightarrow {}^2E$ transition is observed, which can be interpreted by the acentric $Cr^{3+}$ site with very low symmetry (point group $C_1$), thus the Laporte selection rule is largely lifted, which results in high emission possibility for all absorption bands. For doublet terms ($^2G$), the electronic-dipole transition probability is further increased by spin-orbit interaction with spin-allowed quartet terms, as observed in DRS and PLE spectra. The Fano antiresonance features and $^4A_2 \rightarrow {}^2E$ absorption become very strong and sharp at low temperatures, as shown in Fig. 2 (cyan lines). A comparison of spectroscopic data for two phosphors is given in Table 2. Stokes shift associated with the electron-phonon coupling in the $^4T_2$ level is determined to be 2938 cm$^{-1}$ for AWO:Cr and 2888 cm$^{-1}$ for SWO:Cr, respectively. Generally, a larger Stokes shift is accompanied by lower activation energy for thermal quenching of $Cr^{3+}$ luminescence.

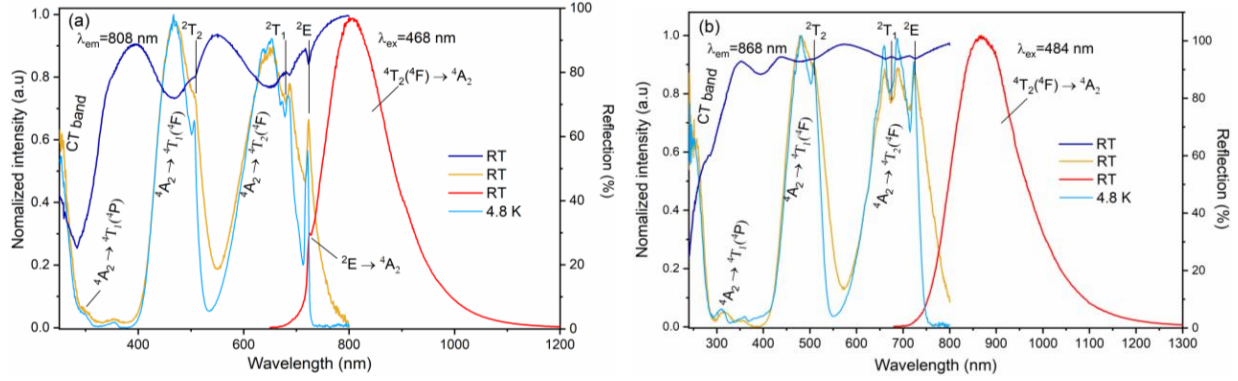

**Fig. 2.** Excitation (orange line), emission (red line), and diffuse reflection spectra of (navy line) AWO:0.04Cr and (b) SWO:0.06Cr phosphors at room temperature (cyan lines are excitation spectra recorded at 4.8 K).

In general, optical spectra of $Cr^{3+}$ can be well described in terms of three crystal field parameters, i.e., crystal field strength ($D_q$), $B$ and $C$ Racah parameters, which can be easily evaluated from the energetic positions of absorption bands detected in PLE spectra of $Cr^{3+}$ ions with following relations given below [43]:

$$D_q = E(^4A_2 \rightarrow {}^4T_2)/10 \tag{1}$$

where $E(^4A_2 \rightarrow {}^4T_2)$ is the energy gap of $^4T_2$ absorption or excitation band.

$$\frac{D_q}{B} = \frac{15(x-8)}{(x)^2 - 10x}, \quad \frac{\Delta E}{D_q} = x \tag{2}$$

$$E(^2E) = 3.05C + 7.9B - 1.8B^2/D_q \tag{3}$$

where $\Delta E$ is the energy difference between $^4T_1(^4F)$ and $^4T_2$ ($^4F$) levels, and $E(^2E)$ is the energy of the $^2E$ level. A comparison of the energetic position of energy levels of $Cr^{3+}$ in AWO and SWO compounds and calculated results of $D_q$, $B$, and $C$ are tabulated in Table 2. The energy level schemes of $Cr^{3+}$ ($3d^4$) ion in an octahedral crystal field are generally described by the well-known Tanabe-Sugano diagram, as shown in Fig. 3a. The smaller value of Dq/B in SWO suggests it to be a weaker CF, which permits a broader and large red-shift in NIR PL spectrum. The nephelauxetic effect, which refers to a decrease in Racah interelectronic repulsion with respect to the value of the free ion, can be expressed by nephelauxetic ratio $\beta=B/B_0$ (where $B_0$ is the free ion value, $B_0$ ($Cr^{3+}$)=918 cm$^{-1}$). The lower value of the $B$ Racah parameter (603 cm$^{-1}$) for AWO in comparison with SWO (645 cm$^{-1}$) indicates a higher degree of covalency in AWO because of significantly shorter Al-O distances and more covalent Al-O bonds. A new parameter $\beta_1 = \sqrt{(B/B_0)^2 + (C/C_0)^2}$ ($B_0$=918 cm$^{-1}$ and $C_0$=3850 cm$^{-1}$ are the Racah parameters of Cr$^{3+}$ in a free state) introduced by Brik and Srivastava [44] can be applied to quantitatively describes the nephelauxetic effect with better accuracy in the spectroscopy of Cr$^{3+}$ ions. The estimated $\beta_1$ values are 1.042 for AWO:0.004Cr and 1.053 for SWO:0.006Cr. The energy level of $^2E$ is then predictable with a linear relationship: $E(^2E)$=3382.80+10021.47$\beta_1$±σ(σ=362 cm$^{-1}$) [45], where σ represents for root-mean square deviation of the data points from the linear fit. With the use of the obtained values, the positions of $^2E$ level for Cr$^{3+}$ in WAO and SWO lines in the range of 13461-14184 and 13578-14302 cm$^{-1}$, respectively, are in good accordance with the experimental data.

**Table 2.** Data of observed energy levels of Cr$^{3+}$, FWHM, Stokes shift, calculated crystal field strength, Racah parameters and β-covalency for AWO:0.04Cr and SWO:0.06Cr.

| Parameter | AWO:Cr | SWO:Cr |
| --- | --- | --- |
| $^4T_1(^4P)$ cm$^{-1}$ | 33333 (300 nm) | 31646 (316 nm) |
| $^4T_1(^4F)$ cm$^{-1}$ | 21368 (468 nm) | 20661 (484 nm) |
| $^2T_2(^2G)$ cm$^{-1}$ | 19802 (505 nm) | 19685 (508 nm) |
| $^2T_1(^2G)$ cm$^{-1}$ | 14706 (680 nm) | 14815 (675 nm) |
| $^2E(^2G)$ cm$^{-1}$ | 13831 (723 nm) | 13793 (725 nm) |
| $^4T_2(^4F)$ cm$^{-1}$ | 15314 (653 nm) | 14409 (694 nm) |
| Maxima (nm) | 12376 (808 nm) | 11521 (868 nm) |
| FWHM (nm) | 140 | 164 |
| $Dq$ cm$^{-1}$ | 1531 | 1441 |
| $B$ cm$^{-1}$ | 603 | 645 |
| $C$ cm$^{-1}$ | 3113 | 3022 |
| $Dq/B$ | 2.54 | 2.23 |
| $\beta=B/B_0$ | 0.67 | 0.70 |
| Stokes-Shift (cm$^{-1}$) | 2938 | 2888 |

As shown in Fig. 3b and c, both AWO:xCr and SWO:xCr (x=1-8%) phosphors show similar broadband emission and slight shifts toward to longer wavelength with increasing Cr$^{3+}$ concentration. The optimal concentration (insert of Fig.3b, c) of Cr$^{3+}$ is determined to be 4% and 6% for AWO and SWO compounds, respectively. The small red shift can be ascribed to the re-absorption of Cr$^{3+}$. The concentration quenching can be ascribed to the energy transfer between the adjacent Cr$^{3+}$ ions by the ways of multipole-multipole interaction, exchange interaction, or re-absorption [17].

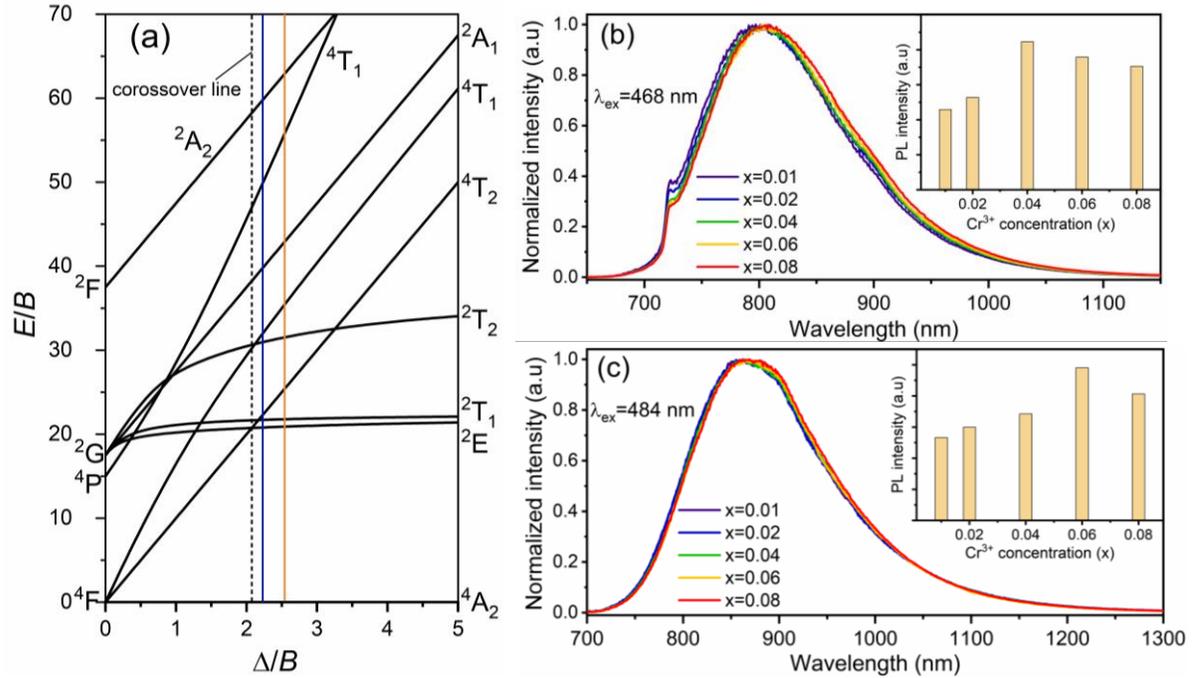

**Fig. 3.** (a) Tanabe-Sugano diagram for the $d^3$ electronic configuration in the octahedral symmetry. Δ/B values for AWO: (blue solid line) and SWO:Cr (orange solid line) are indicated. PL spectra of (b) AWO:xCr and (c) SWO:xCr (x=1-8%) phosphors; insert is PL intensity as a function of the $Cr^{3+}$ concentrations.

### 3.3 Temperature-dependent PL and decay of AWO:0.04Cr and SWO:0.06Cr

In weak crystal fields, the lowest excited state of $Cr^{3+}$ is $^4T_2$, which is derived from the $3d$ electron configuration $t_{2g}^2 e_g$, where the $e_g$ orbital points along the axes of the octahedron directly towards the ligands, thus the $^4T_2 \rightarrow ^4A_2$ transition is therefore strongly coupled to the lattice and tends to suffer from strong electron-phonon coupling effect. The larger Stokes shift in SWO:Cr than that of AWO:Cr implies stronger thermal quenching since the Stokes shift is associated with electron-phonon coupling in the $^4T_2$ state. For better understanding the effect of electron-phonon coupling of $Cr^{3+}$ in studied compounds, the temperature-dependent PL measurements were carried out. The PL spectra of AWO:0.04Cr and SWO:0.06Cr, measured as a function of temperature, are presented in Fig. 4 and Fig. 5. The shape of the spectrum (Fig. 4a), with sharp line and broadband at lower temperatures, is assigned to the spin-forbidden $^2E \rightarrow ^4A_2$ transition superimposed on the spin-allowed $^4T_2 \rightarrow ^4A_2$ transition of $Cr^{3+}$. With the temperature increasing in 4.8-250 K, the PL intensity monotonically decreases and the broad band exhibit a small blue shift (Fig. 4d). This implies the minima of $^2E$ and $^4T_2$ excited state is separated, in agreement with the intermediate crystal field. Interestingly, an abrupt change in the PL intensity was detected at 275 K. The maximum of PL at 275 K is 2 times more intense than that at 250 K, then decreases accompanying red-shift with increasing temperature. The promising change can be mostly ascribed to temperature-induced structural phase transformation from monoclinic (low temperature) with the space group $P2_1$ ($P2_1/n$) to orthorhombic (high temperature) with the space group $Pbcn$ ($D_{2h}^{14}$). Similar phase transition behavior in $Al_2W_3O_{12}$ single crystal was identified earlier by Hanuza et.al [46]. In their report, temperature driven phase transition identified by IR and Raman scattering methods occurs at ~210 K, which is lower than the value of 275 K in the present work analyzed using temperature dependent PL spectra. The decay lifetime of $Cr^{3+}$ luminescence increases abruptly at 275 K, which is indicative of temperature-induced phase transition (Fig. 4e, f). All decay profiles for $^4T_2 \rightarrow {}^4A_2$ transition display a non-exponential feature and thus can be adopted as an equation $\tau = \frac{\int t \cdot I(t)dt}{\int I(t)dt}$. The average decay time calculated becomes shorter from 73.3 to 24.2 μs with the increase in the temperature except for that at 275 K. The abrupt changes in photoluminescence and decay behaviors are clearly observed at 275 K, in the process of heating and cooling, thereby revealing the reversible phase transition in AWO phosphor.

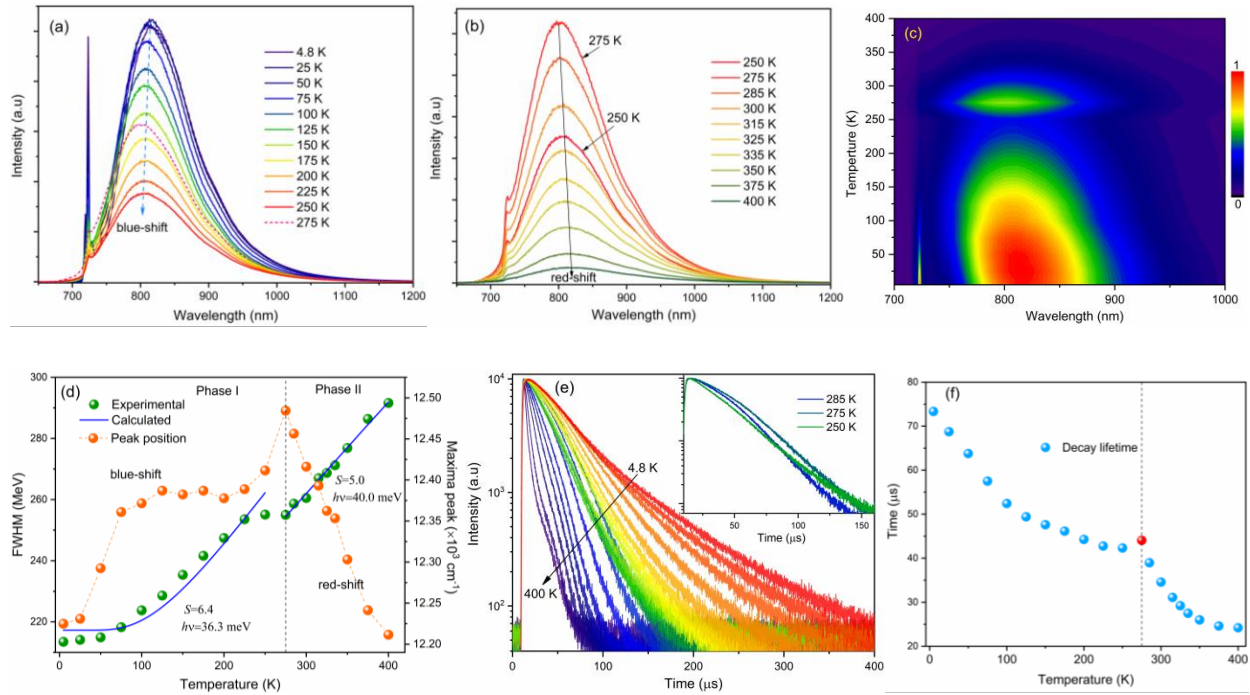

**Fig. 4.** Temperature-dependent PL emission spectra of AWO:0.04Cr measured in the 4.5-275 K (a) and 250-400 K range (b). (c) the corresponding wavelength-temperature contour plot. (d) maximum peak position and fitting of the FWHM ($^4T_2$ band) as a function of temperature. (e) Temperature dependent decay curves (e) and (f) variation of the average lifetime (monitored at 808 nm upon excitation at 468 nm).

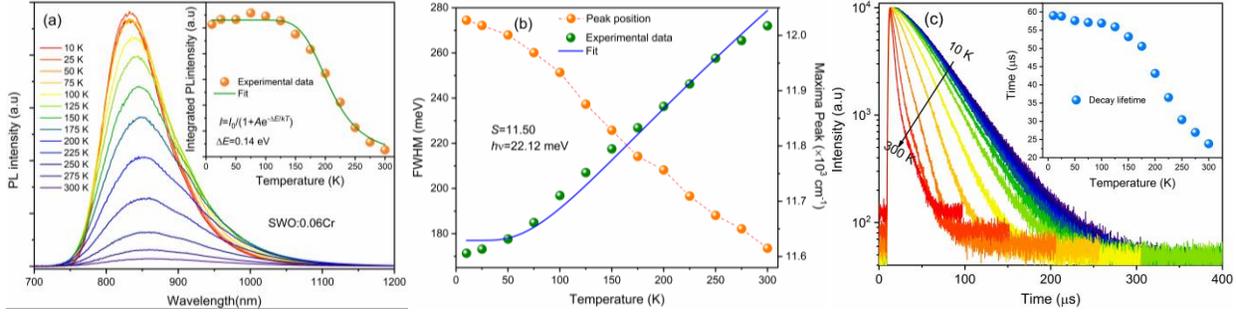

**Fig. 5.** (a) Temperature dependence of PL emission spectrum of SWO:0.06Cr in the 4.8-300 K range; The inset showing integrated PL intensity vs. temperature and Arrhenius eq. fit; The calculated deactivation energy is given in the figure. (b) Temperature dependence of maximum peak position and fit of the FWHM as a function of temperature. (c) Temperature dependent decay curves and variation of lifetimes (insert) (monitored at maximum peak upon excitation at 484 nm).

Compared to AWO:0.04Cr, the SWO:0.06Cr shows a stronger quenching of fluorescence and decay time with temperature (Fig. 5a, c). The PL decay time gradually decreases from 59.0 to 23.8 μs. The temperature dependence of the PL intensity can be well fitted by the Arrhenius formula $I(t) = I_0 \big/ \left(1 + Ae^{\Delta E/kT}\right)$, and the activation energy defined as energy separation between minima of $^4T_2$ state and its crossover point to $^4A_2$ ground state is calculated to be $\Delta E$= 0.14 eV, suggesting a stronger electron-phonon (EP) coupling of the $^4T_2 \rightarrow {}^4A_2$ emission from $Cr^{3+}$ in SWO. The EP coupling can be determined by the vibration of FWHM and spectral shift (Fig. 5b) as a function of temperature. The maxima of the $^4T_2 \rightarrow {}^4A_2$ band of SWO:0.06Cr shift from 830 nm at 4.8 K to 860 nm at 300 K. The FWHM as a function of temperature can be described by Boltzmann distribution function given by [47],

$$FWHM = 2.36 \times h\nu \times \sqrt{S} \times \sqrt{\coth\left(\frac{h\nu}{2kT}\right)} \qquad (4)$$

where $h\nu$ is the effective phonon energy that interacts with the electronic transitions, $T$ is the Kelvin temperature, $S$ is the dimensionless Huang-Rhys parameter and $k$ is the Boltzmann constant ($0.8617×10^{-5}$ eV). The best fit to eq.4 for SWO gives the parameter $S$ of 11.5, which is about twice as large as that for AWO, implying stronger EP coupling in SWO. The weaker electron-phonon coupling strength in AWO explains the weaker thermal quenching for the $Cr^{3+}$ emission in AWO in comparison with SWO. On the other hand, the temperature quenching and broadening behavior of the FWHM can be explained by the use of the configurational coordinate diagram. Temperature increases cause the phonon vibration of greater amplitude and thus the effect of electron-phonon interaction is dominant, resulting in broader vibrational energy distribution and a broader PL emission band. Meanwhile, the luminescence of $Cr^{3+}$ is quenched by the thermally activated crossing of the $^4T_2$ excited state and $^4A_2$ ground state.

### 3.4 Raman spectroscopy of AWO:Cr under high pressure

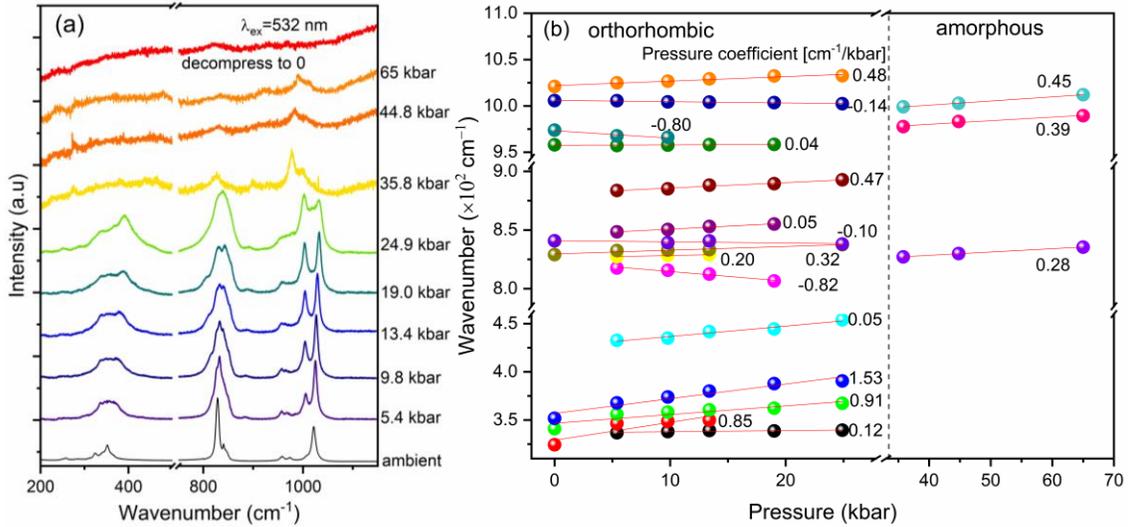

**Fig. 6.** Raman spectra (a) of AWO:0.04Cr excited with the 532 nm laser line at room temperature as a function of pressure and (b) the frequency shifts as a function of pressure. Solid lines are linear fits to the data.

To gain insight into the phase transformation of AWO compound Raman spectra of AWO:0.04Cr as a function of pressure up to 65 kbar, measured at room temperature, were presented in Fig. 6a. At room temperature and ambient pressure, the orthorhombic phase (space group $D_{2h}^{14}=Pbcn$) of AWO contains 68 atoms per unite cell. The structure consists of a corner-shared network of $WO_4$ tetrahedra and $AlO_6$ octahedra, as described in Section 3. $Al^{3+}$ ions occupy sites of $C_1$ symmetry and the $WO_4^{2-}$ tetrahedra form two crystallographically distinct sets of $C_1$ and $C_2$ symmetry. The factor group analysis leads to 204 degree of freedom ($k=0$), which are described by following irreducible representations: $25A_g+26B_{1g}+25B_{2g}+26B_{3g}+25A_u+26B_{1u}+25B_{2u}+26B_{3u}$ [48]. Among these, only $A_g$, $B_{1g}$, $B_{2g}$, and $B_{3g}$ modes are Raman active. shows the Raman spectrum of AWO:0.04Cr at room temperature. The observed Raman bands (Fig. 7a) are analyzed in a similar way as reported earlier in AWO single crystals[46,49]. The modes in the frequency range 900-1500 cm$^{-1}$ could be assigned to the symmetric stretching and those in the range 800-900 cm$^{-1}$ to the antisymmetric stretching modes of WO$_4$ tetrahedral. The prominent peaks observed between 300 and 400 cm$^{-1}$ are identified as antisymmetric bending and symmetric bending modes of WO$_4$.

Representative Raman spectra of AWO:Cr under high pressures and the frequency shifts are shown in Fig. 6a, b, respectively. All the Raman peaks observed at ambient pressure are relatively strong and sharp. The shift of these peaks can be well fitted with a linear pressure dependence up to 24.9 kabr. With the increase of pressure, most Raman bands shift to the higher frequencies and broaden with diminished intensity. It is worth noting that four modes exhibit negative slopes, which is typically found in materials displaying negative thermal quenching properties [50]. Above 24.9 kbar, significant spectral changes occur, as most peaks disappear and only three new peaks with very low intensity can be identified, indicating the onset of a structural phase transformation. After decompression, no Raman modes are discernible, which implies that the structure most likely transforms into an amorphous. A similar phenomenon was observed in Al$_2$W$_3$O$_{12}$ utilizing high-pressure synchrotron X-ray powder diffraction [51].

### 3.5 Performance of the fabricated NIR pc-LED device

A high-power NIR pc-LED (insert of Fig. 7a) was fabricated by coating AWO:0.04Cr on a commercial 470 nm LED chip as part of our assessment as NIR light source material. The electroluminescence (EL) spectra, optical powers, and conversion efficiencies of the device to the driving current ($I$) are presented in Fig. 7a,b. The weak emission band ~470 nm arises from the blue chip, while the strong broad NIR emission band in 700-1000 nm is ascribed to AWO:0.04Cr$^{3+}$. The NIR output power of 17.1 mW was obtained with a driving current at 320 mA. NIR photoelectric efficiency declines from 4.0% to 1.9% as the current increases from 20 to 320 mA. To be competitive with other NIR light sources, the performance of NIR pc-LED should be further optimized.

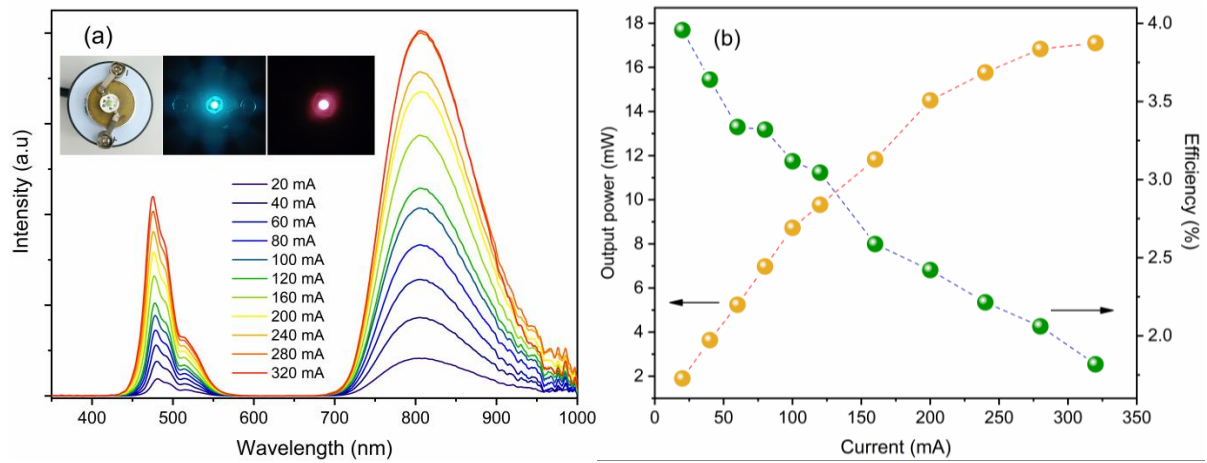

**Fig. 7.** (a) The emission spectra of pc-NIR LED fabricated with AWO:0.04Cr$^{3+}$ phosphor on a blue chip (λ=470 nm). (b) The driven current dependence of output powers and conversion efficiencies.

## 4. Conclusions

In conclusion, we propose Cr$^{3+}$-activated Al$_2$(WO$_4$)$_3$ and Sc$_2$(WO$_4$)$_3$ phosphors as potential NIR light source materials for NIR pc-LED. Upon blue excitation, these compounds permit broadband NIR emission stems primarily from $^4T_2 \rightarrow {^4A_2}$ transition in the range of 670-1200 nm (maxima ~808 nm, FWHM ~140 nm) for AWO:Cr and of 700-1300 nm (maxima ~870 nm, FWHM ~164 nm) for SWO:Cr. The crystal field parameters and nephelauxetic effect of octahedrally coordinated Cr$^{3+}$ in both compounds have been analyzed based on spectroscopic results. The electron-phonon coupling parameter $S$ for SWO:Cr was determined to be 11.5, twice as large as that for AWO:Cr, which explains its stronger thermal quenching and lower quenching temperature. An abrupt change observed at 275 K in temperature-dependent luminescence spectra and decay lifetime of AWO:Cr is associated with temperature-driven phase transition from monoclinic to orthorhombic. Pressure-induced amorphization above 25 kbar was confirmed for AWO:Cr materials through high pressure evolution of Raman spectra. A high-power pc-NIR LED, fabricated by coating AWO:0.04Cr on a commercial 470 nm LED chip, shows good performance with an output power of 17.1 mW driven by a current of 320 mA, which reveals the potential application of as-prepared materials in NIR light sources.

## Acknowledgement

This work was financially supported by National Natural Science Foundation of China (12004062), the Science and Technology Research Program of Chongqing Municipal Education Commission (Grant No. KJQN202100615, KJZD-M202000601), and Natural Science Foundation of Chongqing (cstc2021jcyj-msxmX0277). Y. Y. Zhang thanks the supports from the China Postdoctoral Science Foundation (2021M702562), Guangdong Basic and Applied Basic Research Foundation (21201910240000723), and the Research Start-up Funds of DGUT (211135307). A. S. acknowledge the support of National Science Centre, Poland, grant SHENG 2 number: 2021/40/Q/ST5/00336.

# References

[1] K. B. Beć, J. Grabska and C.W. HucK, Near-infrared spectroscopy in bio-applications, Molecules. 25 (2020) 2948.

[2] F. Vatansever, M.R. Hamblin, Far infrared radiation (FIR): its biological effects and medical applications, Photonics Lasers med. 1 (2012) 255–266.

[3] D. Hayashi, A.M. van Dongen, J. Boerekamp, S. Spoor, G. Lucassen, J. Schleipen, A broadband LED source in visible to short-wave-infrared wavelengths for spectral tumor diagnostics, Appl. Phys. Lett. 110 (2017) 233701.

[4] B.M. Nicolai, K. Beullens, E. Bobelyn, A. Peirs, W. Saeys, K.I. Theron, J. Lammertyn, Nondestructive measurement of fruit and vegetable quality by means of NIR spectroscopy: a review, Postharvest Biol. Technol. 46 (2007) 99–118.

[5] Y.H. Chien, Y.L. Chou, S.W. Wang, S.T. Hung, M.C. Liau, Y.J. Chao, C.H. Su, C.S. Yeh, Near-infrared light photocontrolled targeting, bioimaging, and chemotherapy with caged upconversion nanoparticles in vitro and in vivo, ACS Nano. 7 (2013) 8516–8528.

[6] J. Coaton, Modern tungsten-halogen-lamp technology, Proc. Inst. Electr. Eng., 117(1970) 1953-1959.

[7] V. Rajendran, H. Chang, Ru-Shi Liu, Recent progress on broadband near-infrared phosphors converted light emitting diodes for future miniature spectrometers, Opt. Mater. X 1( 2019) 100011.

[8] E.F. Schubert, Light-Emitting Diodes, Cambridge University Press, Cambridge, 2003.

[9] B. Henderson, R.H. Bartram, Crystal-Field Engineering of Solid-State Laser Materials, Cambridge University Press, 2000.

[10] A.B. Budgor, L. Esterowitz, and L.G. DeShazer, Tunable solid-state laser II, Series in Optical Sciences, Vol. 52, Spring-Verlag, Berlin, Heidelberg, 1986.

[11] S.A. Payne, L.L. Chase, L.K. Smith, W.L. Kway, and H.W. Newkirk, Laser performance of LiSrAlF$_6$:Cr$^{3+}$, J. App. Phys. 66 (1989) 1051.

[12] Q. Shao, H. Ding, L. Yao, J. Xu, C. Liang, J. Jiang, Photoluminescence properties of a ScBO$_3$:Cr$^{3+}$ phosphor and its applications for broadband near-infrared LEDs, RSC Adv. 8 (2018) 12035−12042;

[13] B. Malysa, A. Meijerink, T. Jüstel, Temperature dependent luminescence Cr$^{3+}$-doped GdAl$_3$(BO$_3$)$_4$ and YAl$_3$(BO$_3$)$_4$, J. Lumin. 171 (2016) 246–253.

[14] H. Zeng, T. Zhou, L. Wang, R. Xie, Two-site occupation for exploring ultra-broadband near-infrared phosphor-double-perovskite La$_2$MgZrO$_6$:Cr$^{3+}$, Chem. Mater. 31 (2019) 5245−5253.

[15] B. Malysa, A. Meijerink, W. Wu, T. Juestel, On the influence of calcium substitution to the optical properties of Cr$^{3+}$ doped SrSc$_2$O$_4$, J. Lumin. 190 (2017) 234−241.

[16] Z.W. Jia, C.X. Yuan, Y.F. Liu, X.J. Wang, P. Sun, L. Wang, H. C. Jiang, J. Jiang, Strategies to approach high performance in Cr$^{3+}$-doped phosphors for high-power NIR-LED light sources, Light Sci. Appl. 9 (2020) 2047-7538.

[17] M. Mao, T. Zhou, H. Zeng, L. Wang, F. Huang, X. Tang and R. Xie, Broadband near-infrared (NIR) emission realized by the crystal-field engineering of Y$_{3-x}$Ca$_x$Al$_{5-x}$Si$_x$O$_{12}$:Cr$^{3+}$ (x=0-2.0) garnet phosphors, J. Mater. Chem. C 8 (2020) 981-1988.

[18] B. Malysa, A. Meijerink, T. Juestel, Temperature dependent Cr$^{3+}$ photoluminescence in garnets of the type X$_3$Sc$_2$Ga$_3$O$_{12}$ (X = Lu,Y, Gd, La), J. Lumin. 202 (2018) 523-531.

[19] L. Zhang, S. Zhang, Z. Hao, X. Zhang, G. Pan, Y. Luo, H. Wu, J. Zhang, A high efficiency broad-band near-infrared Ca$_2$LuZr$_2$Al$_3$O$_{12}$:Cr$^{3+}$ garnet phosphor for blue LED chips, J. Mater. Chem. C 6 (2018) 4967-4976.

[20] L. L. Zhang, D. D. Wang, Z. D. Hao, X. Zhang, G. H. Pan, H. J. Wu, J. H. Zhang. Cr$^{3+}$-Doped Broadband NIR Garnet Phosphor with Enhanced Luminescence and its Application in NIR Spectroscopy, Adv. Opt. Mater. 7 (2019) 1900185.

[21] S. He, L.L. Zhang, H. Wu, H.J. Wu, G.H. Pan, Z.D. Hao, X. Zhang, L.G. Zhang, H. Zhang, J.H. Zhang. Efficient Super Broadband NIR Ca$_2$LuZr$_2$Al$_3$O$_{12}$:Cr$^{3+}$,Yb$^{3+}$ Garnet Phosphor for pc-LED Light Source toward NIR Spectroscopy Applications, Adv. Opt. mater. 2020, 1901684, 1-7.

[22] E.T. Basore, W.G. Xiao, X.F. Liu, J.H. Wu, J.R. Qiu, Broadband near-infrared garnet phosphors with near-unity internal quantum efficiency, Adv. Optical. Mater. 8 (2020) 2000296.

[23] W.D. Nie, L.Q. Yao, G. Chen, S.H. Wu, Z.J. Liao, L. Han, X.Y. Ye, A novel Cr$^{3+}$-doped Lu$_2$CaMg$_2$Si$_3$O$_{12}$ garnet phosphors with broadband emission for near-infrared applications, Dalton Trans. 50 (2021) 8446-8456.

[24] D. Huang, S. Liang, D. Chen, J. Hu, K. Xu, H. Zhu, An efficient garnet-structured $Na_3Al_2Li_3F_{12}:Cr^{3+}$ phosphor with excellent photoluminescence thermal stability for near-infrared LEDs, Chem. Eng. J. 426 (2021) 131332.

[25] X. Zou, X. Wang, H. Zhang, Y. Kang, X. Yang, X. Zhang, M.S. Molokeev, A highly efficient and suitable spectral profile $Cr^{3+}$-doped garnet near-infrared emitting phosphor for regulating photomorphogenesis of plants, Chem. Eng. J. 428 (2022) 132003.

[26] T. Lang, M. Cai, S. Fang, T. Han, S. He, Q. Wang, G. Ge, J. Wang, C. Guo, L. Peng, S. Cao, B. Liu, V.I. Korepanov, A.N. Yakovlev, J. Qiu, Trade-off Lattice Site Occupancy engineering strategy for near-infrared phosphors with ultrabroad and tunable emission, Adv. Optical. Mater. 10 (2022) 2101633.

[27] J. Qiao, S. Zhang, X. Zhou, W. Chen, R. Gautier, Z. Xia. Near-infrared light-emitting diodes utilizing a europium-activated calcium oxide phosphor with external quantum efficiency of up to 54.7%, Adv. Mater. 34 (2022) 2201887.

[28] Z. Yang, Y. Zhao, Y. Zhou, J. Qiao, Y. Chuang, M. S. Molokeev, Z. Xia. Giant red-shifted emission in $(Sr,Ba)Y_2O_4:Eu^{2+}$ phosphor toward broadband near-Infrared luminescence. Adv. Fun. Mater. 32 (2022) 2103927.

[29] B. Su, M. Li, E. Song, Z. Xia. $Sb^{3+}$-doping in cesium zinc halides single crystals enabling high-efficiency near-infrared emission, Adv. Fun. Mater. 31 (2021) 2105316.

[30] Dongjie Liu, G. Li, P. Dang, Q. Zhang, Y. Wei, L. Qiu, M. S. Molokeev, H. Lian, M. Shang, J. Lin. Highly efficient $Fe^{3+}$-doped $A_2BB'O_6$ ($A=Sr^{2+}, Ca^{2+}$; $B,B' = In^{3+}, Sb^{5+}, Sn^{4+}$) broadband near-infrared-emitting phosphors for spectroscopic analysis, Light Sci. Appl. 11 (2022) 112.

[31] J. S. O. Evans, T.A. Mary, A.W. Sleight, Negative thermal expansion in a large molybdate and tungstate family, J. Solid State Chem. 133 (1997) 580-583.

[32] J. S. O. Evans, T.A. Mary, A.W. Sleight, Negative Thermal Expansion in $Sc_2(WO_4)_3$, J. Solid state Chem. 137 (1998) 148-160.

[33] N. Imanaka, T. Ueda, Y. Okazaki, S. Tamura, G. Adachi, Trivalent Ion Conduction in Molybdates Having $Sc_2(WO_4)_3$-Type Structure, Chem. Mater. 12 (2000) 1910–1913.

[34] K. Petermann, P. Mitzscherlich, Spectroscopic and laser properties of $Cr^{3+}$-doped $Al_2(WO_4)_3$ and $Sc_2(WO_4)_3$, J. Quantum Electron. 23 (1987) 1122-1126.

[35] G. Wang, Z. Lin, L. Zhang, Y. Huang, G. Wang, Spectral characterization and energy levels of $Cr^{3+}:Sc_2(MoO_4)_3$ crystal, J. Lumin. 129 (2009) 1398-1400.

[36] D. Ivanova, V. Nikolov, R. Todorov, Single crystals growth and absorption spectra of $Cr^{3+}$-doped $Al_{2-x}In_x(WO_4)_3$ solid solutions, J. Cryst. Growth 311 (2009) 3428-3434.

[37] S.C. Abrahams and J.L. Bernstein, Crystal Structure of the Transition-Metal Molybdates and Tungstates. II.Diamagnetic $Sc_2(WO_4)_3$, J. Chem. Phys. 45 (1966), 2745-2752.

[38] M.D. Sturge, H.J. Guggenheim, M.H.L. Pryce, Antiresonance in the optical spectra of transition-metal ions in crystals, Phys. Rev. B 2(1970) 2459–2471.

[39] W. Kolbe, K. Petermann and G. Huber, Broadband emission and laser Action of $Cr^{3+}$ Doped Zinc tungstate at 1 um wavelength, IEEE J. Quantum Electron. 21(1986) 1596-1599.

[40] M. Wildner, A. Beran, F. Koller, Spectroscopic characterisation and crystal field calculations of varicoloured kyanites from Loliondo, Tanzania. Miner. Petrol. 107 (2013) 289-310.

[41] O. Maalej, O. Taktak, B. Boulard, S. Kammoun, Study with analytical equations of absorption spectra containing interference dips in fluoride Glasses Doped with $Cr^{3+}$, J. Phys. Chem. B 120 (2016) 7538-7545.

[42] L. Li, Y. Yu, G. Wang, L. Zhang, Z. Lin, Crystal growth, spectral properties and crystal field analysis of $Cr^{3+}$:$MgWO_4$, Cryst. Eng. Comm. 15 (2013) 6083-6089.

[43] Y. Tanabe and S. Sugano, On the Absorption Spectra of Complex Ions II, J. Phys. Soc. Jpn. 9 (1954) 766-779.

[44] A.M. Srivatava, M.G. Brik, Crystal field studies of the $Mn^{4+}$ energy levels in the perovskite $LaAlO_3$, Opt. Mater. 35(2013)1544-1548.

[45] M.G. Brik, S.J. Camardello, A. M. Srivastava, Influence of Covalency on the $Mn^{4+}$ $^2E_g \rightarrow ^4A_{2g}$ Emission Energy in Crystals, ECS J. Solid State Sci. Technol. 4 (2015) R39-R43.

[46] J. Hanuza, M. Mączka, K. Hermanowicz, M. Andruszkiewicz, A. Pietraszko, W. Stręk, P. Dereń, The structure and spectroscopic properties of $Al_{2-x}Cr_x(WO_4)_3$ crystals in orthorhombic and monoclinic Phases, J. Sol. State Chem. 105 (1993) 49-69.

[47] F.Y. Zhao, Z. Song, J. Zhao and Q.L. Liu, Double perovskite $Cs_2AgInCl_6$:$Cr^{3+}$: broadband and near-infrared luminescent materials, Inorg. Chem. Front. 6 (2019) 3621.

[48] M. Maczka, W. Paragussu, A.G. Souza, P.T.C. Freire, J.M. Filho, F.E.A Melo and J. Hanuza, High-pressure Raman study of $Al_2(WO_4)_3$, J. Solid State Chem. 177 (2004) 2002–2006

[49] W. Paragussu, M. Maczka, A.G. Souza, P.T.C. Freire, F.E.A Melo, J.M. Filho and J. Hanuza, A comparative study of negative thermal expansion materials $Sc_2(MoO_4)_3$ and $Al_2(WO_4)_3$ crystals. Vib. Spectrosc. 44 (2007) 69-77.

[50] T. R. Ravindran, A. K. Arora, T. A. Mary, High Pressure Behavior of $ZrW_2O_8$: Grüneisen Parameter and Thermal Properties, Phys. Rev. Lett. 84 (2000) 3879.

[51] T. Varga, A.P. Wilkinson, C. Lind, W.A. Bassett, C. Zha, High pressure synchrotron X-ray powder diffraction study of $Sc_2Mo_3O_{12}$ and $Al_2W_3O_{12}$, J. Phys. Condens. Matter. 17 (2005) 4271-4283.